\documentclass{article}
\usepackage{bagrow}

\usepackage{xr-hyper}
\usepackage[pdftex, pdftitle={Article},colorlinks, pdfauthor={Author}]{hyperref} %
\usepackage[nameinlink,capitalise]{cleveref}
\definecolor{darkgray}{gray}{0.35}
\hypersetup{colorlinks={true},linkcolor={black},citecolor=ForestGreen}

 \usepackage{filecontents}
\usepackage{pdfpages}

\usepackage{xr}
\makeatletter
\newcommand*{\addFileDependency}[1]{
  \typeout{(#1)}
  \@addtofilelist{#1}
  \IfFileExists{#1}{}{\typeout{No file #1.}}
}
\makeatother

\newcommand*{\myexternaldocument}[1]{
    \externaldocument[si:]{#1}
    \addFileDependency{#1.tex}
    \addFileDependency{#1.aux}
}
\myexternaldocument{SI}
\crefname{section}{Sec.}{Secs.}
\Crefname{section}{Section}{Sections}
\crefname{figure}{Fig.}{Figs.}
\Crefname{figure}{Figure}{Figures}
\crefname{equation}{Eq.}{Eqs.}
\Crefname{equation}{Equation}{Equations}

\makeatletter

\renewcommand*{\p@subsection}{}

\renewcommand*{\p@subsubsection}{}
\makeatother

\usepackage{comment,soul}
\definecolor{reddish}{HTML}{FBB4AE}
\definecolor{blueish}{HTML}{B3CDE3}
\definecolor{magentish}{HTML}{FF00AA}
\definecolor{greenish}{HTML}{a1d99b}
\definecolor{orangeish}{HTML}{ffdead}

\newcommand{\avg}[1]{\ensuremath{\left<{#1}\right>}}
\newcommand{\mat}[1]{\ensuremath{\mathbf{#1}}}
\newcommand{\Emat}{\ensuremath{\mat{E}}}
\newcommand{\Bmat}{\ensuremath{\mat{B}}}

\newcommand{\Nmat}{\ensuremath{\mat{N}}}
\newcommand{\Ehat}{\ensuremath{\hat{\mat{E}}}}

\title{Recovering lost and absent information in temporal networks}

\author[1,2,*]{James P.~Bagrow}
\author[3,4]{Sune Lehmann}
\affil[1]{Mathematics \& Statistics, University of Vermont, Burlington, VT, United States }
\affil[2]{Vermont Complex Systems Center, University of Vermont, Burlington, VT, United States}
\affil[3]{DTU Compute, Technical University of Denmark, Kgs Lyngby, Denmark}
\affil[4]{Center for Social Data Science, University of Copenhagen, Copenhagen, Denmark}
\affil[*]{\corrauthinfo{james.bagrow@uvm.edu}{bagrow.com}
}

\newcommand{\frob}[1]{\left\lVert#1\right\rVert_F}

\date{July 22, 2021}

\begin{document}

\maketitle

\begin{abstract}\onehalfspacing
The full range of activity in a temporal network is captured in its edge activity data---time series encoding the tie strengths or on-off dynamics of each edge in the network.
However, in many practical applications, edge-level data are unavailable, and the network analyses must rely instead on node activity data which aggregates the edge-activity data and thus is less informative. 
This raises the question: 
Is it possible to use the static network to recover the richer edge activities from the node activities?
Here we show that recovery is possible, often with a surprising degree of accuracy given how much information is lost, and that the recovered data are useful for subsequent network analysis tasks.
Recovery is more difficult when network density increases, either topologically or dynamically, but exploiting dynamical and topological sparsity enables effective solutions to the recovery problem.
We formally characterize the difficulty of the recovery problem both theoretically and empirically, proving the conditions under which recovery errors can be bounded and showing that, even when these conditions are not met, good quality solutions can still be derived.
Effective recovery carries both promise and peril, as it enables deeper scientific study of complex systems but in the context of social systems also raises privacy concerns when social information can be aggregated across multiple data sources.
\end{abstract} 

\section{Introduction}

Temporal networks are of increasing importance as models for systems across across social, biological and technological domains~\cite{holme2012temporal, bassett2017network, li2017fundamental}. 
Their study has been predicated on data and high-resolution, high-velocity data are becoming more common across fields~\cite{holme2015modern}.
In a temporal network, each link turns on and off over time, forming an intricate pattern of dynamic connectivity~\cite{masuda2016guide}.
In many real-world instances---such as social networks captured from smartphones~\cite{deMontjoye2018privacyConscientious}, gene regulatory networks obtained via high-throughput experiments~\cite{natureReviewsGenetics2012}, and brain networks measured through neuroimaging~\cite{Bassett7641,faskowitz2020edge}---these rich dynamics are unavailable.
Instead, we can only know the activation patterns of nodes: the information regarding when a node connects to some other node, but not which specific one.
In some cases, this is by design: to preserve user privacy, a smartphone maker may prevent an installed application from accessing user communication records~\cite{horvat2012one, garcia2017leaking, bagrow2019information}, but allow that application access to, for example, geolocation~\cite{de2013unique} or accelerometer signals~\cite{kroger2019privacy}.
In other cases, it may be too costly or otherwise infeasible to monitor the activities of every edge. 
In many biological systems, we simply lack the experimental tools to directly capture \textit{in situ} edge activities~\cite{natureReviewsGenetics2012,altelaar2013nextGenProteomics,faskowitz2020edge}.

While node activities are generally more available in empirical temporal network data, the loss of information compared to edge activities is currently not understood.
This leads us to ask the following question:
if a researcher loses, or otherwise cannot ascertain, the edge activity data, but does have access to the node activities, how well can they recover the richer edge activity data?
Solely based on nodal activation, at best, this recovery is only possible in a small number of special cases.
However, if the network's static structure is available, e.g. either directly available from a separate measurement or can be inferred via side information, then it may be possible to recover, even if only approximately, the absent edge activity data.
Such side information can take the form of additional experiments or joining across datasets.
For brain networks, for example, tract-tracing~\cite{wedeen2005mapping} or fluorescence microscopy~\cite{livet2007transgenic} studies revealing network connectivity can complement functional neuroimaging~\cite{faskowitz2020edge}, while for social behavior, data brokers~\cite{anthes2014dataBrokers} may be able to gather the structure of user social networks by cross-referencing public and private social activity datasets~\cite{garcia2017leaking,bagrow2019information}.

There is both promise and peril in the recovery of edge activities.
In the case of biological systems, this opens up new possibilities for understanding the fundamental dynamics of living cells.
In the context of social networks, rich data raises privacy concerns.
Data brokers, who may be able to aggregate across both public and nonpublic sources of information, have the potential to circumvent privacy protections and reveal more about individuals than would be possible without such high-resolution data.

In this paper, we make the following contributions.
In \cref{sec:networkrecovery} we define and motivate the problem of information loss in temporal networks and how to recover it.
We apply solution methods to a representative network corpus, %
including studying how well network analysis tasks can be performed on recovered data (\cref{sec:networktasksafterrecovery}) and better understanding both theoretically and empirically the topological and dynamical sources of recovery error (\cref{sec:understandingrecoveryerrors}).
We conclude with a discussion providing context for the recovery problem across different scientific domains in \cref{sec:discussion}, including both the benefits for advancing network measurement and the concerns with protecting privacy raised by this work.

\section{Information loss and recovery in temporal networks}
\label{sec:networkrecovery}

\begin{figure}
    \centering
    \includegraphics[width=\textwidth]{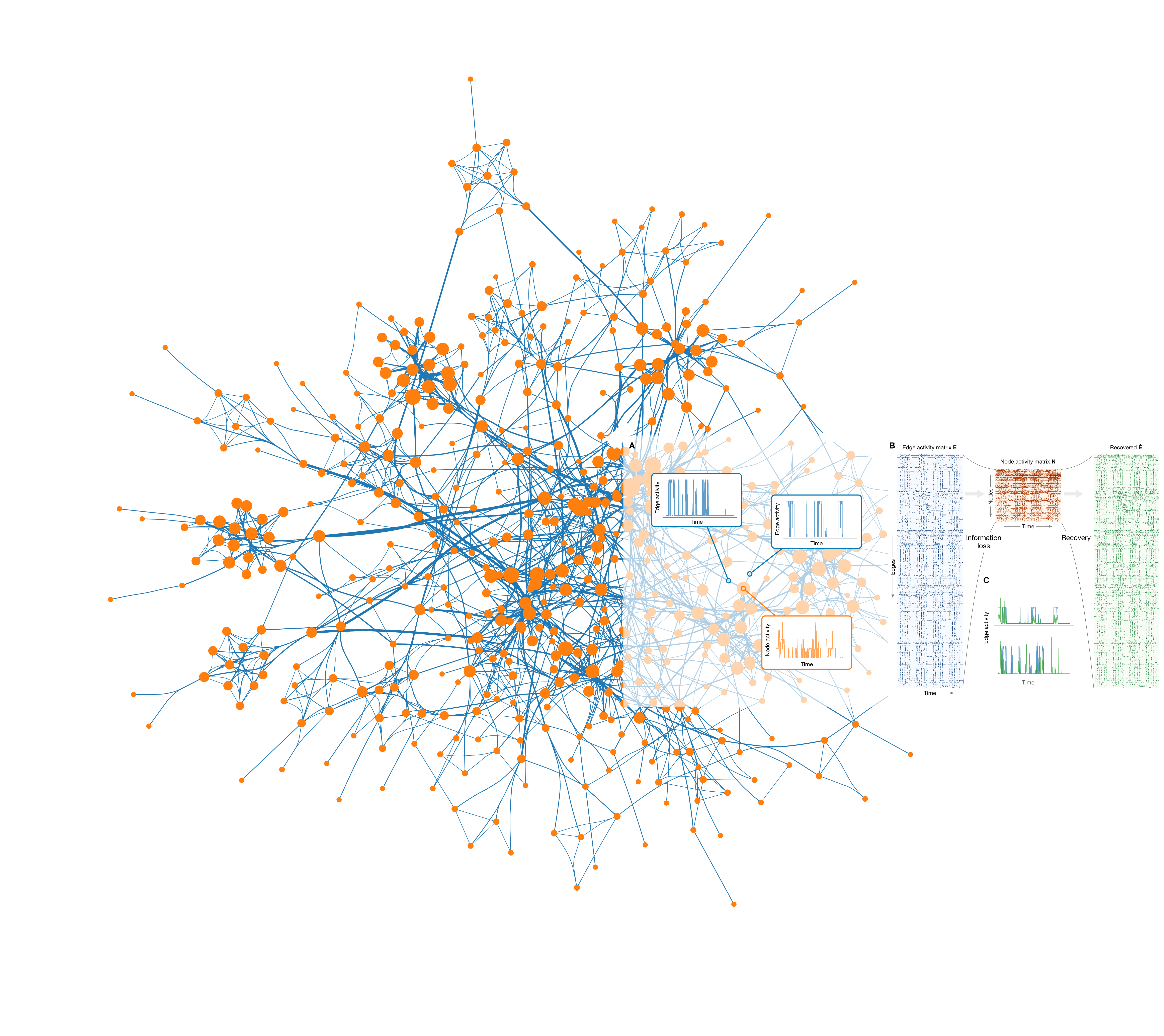}
    \caption{Information loss and recovery in temporal networks.
    \lett{A} Using the Copenhagen Network to illustrate two edge activity time series' and a node's corresponding activity (aggregated for all its incident edges).
    \lett{B} Significant information is lost when only the much smaller node activity matrix is available. 
    (The Copenhagen network has an average degree $\avg{k}\approx 8.8$ meaning there are $\approx 4.4$ edges per node and thus an over four-fold reduction in information when going from $\mat{E}$ to $\mat{N}$)
    However, as the network topology is learned, the edge activity matrix can be recovered, potentially with high accuracy, $\mat{E} \approx \hat{\mat{E}}$.
    \lett{C}
    Comparing the original and recovered edge activities for the two edges highlighted in panel A. 
    }
    \label{fig:cartoon}
\end{figure}

Suppose a researcher is interested in studying a time-evolving, or temporal, network, fully defined through its edge activations.
For many networks, such edge data are not always readily available.
It may be impossible, costly or otherwise impractical to monitor the activities of every edge in the network.
In some cases, one must even observe every pair of nodes, a costly, quadratic operation.
Therefore, in practice one often lacks the ideal edge activity data.
We argue that progress on studying the network's time dynamics can still be made, however, by studying the time series of node activities, when available.
Node activities, which describe when a node is active but not with whom, are less informative but more readily available, as monitoring a singleton is less resource-intensive than monitoring a pair. 
Further, there are fewer singletons that need to be monitored than duos, meaning there will be (unless the network is exceptionally sparse) fewer node time series than edge time series.
Yet, while node activity data are more practical, the loss of information is typically significant.
The amount of information lost depends on network structure, which determines how edge activities are projected down to the node activities. (See     \cref{fig:cartoon}A, illustrated using `Copenhagen', the physical proximity component of the Copenhagen Networks Study dataset~\cite{sapiezynski2019interaction}, a smartphone-derived social network which is part of our corpus of temporal networks; described below). 

We now formalize the information loss and recovery problem.
Consider a network $G=(V,E)$ of $n=\left|V\right|$ nodes (or vertices) and $m = \left|E\right|$ edges (or links).
Let $\mat{E}$ by an $m \times T$ matrix of edge activity data, where $E_{ij,t}$ is the number of interactions between $i$ and $j$ during timestep $t$ (or during a time window $(t,t+\Delta t)$). 
Each row of $\mat{E}$ corresponds to a time series of $T$ values for an edge in the network. 
These time series' encode the temporal activities of the network, when edges are active and at what strength.
A reduced view of the temporal network can be given by the node activity matrix $\mat{N}$.
Here $\mat{N}$ is an $n \times T$ matrix where each row captures the activity of a corresponding node in the network.

The edge and node activity matrices are related to one another via the graph \emph{incidence matrix} $\mat{B}$.
The (unoriented) incidence matrix is an $n \times m$ binary matrix that details which edges are incident on a node.
Using $\mat{B}$, we then have
\begin{equation}
    \mat{B} \mat{E} = \mat{N}.
    \label{eqn:nodeandedgematrices}
\end{equation}
This relationship shows us that the edge activity matrix is more fundamental than the node matrix given the network structure, as $\mat{N}$ can be derived from $\mat{E}$.
\Cref{fig:cartoon}B illustrates this information loss by the reduction in size as $\Emat$ becomes $\Nmat$ (\cref{eqn:nodeandedgematrices}).

Now, suppose we do not have the edge activities but we do have the node activities. 
The question now becomes: can we recover $\Emat$ from $\Nmat$ using the network structure (\cref{fig:cartoon}B,C)?
\Cref{eqn:nodeandedgematrices} is a system of linear equations in the standard form $A x = b$, which can usually be solved by, for example, $x = A^{-1}b$.
However, $\Bmat$ is a rectangular matrix and, except for certain edge cases such as a network with no cycles, \cref{eqn:nodeandedgematrices} is an underdetermined system.
This means there will be either no solutions or an infinite number of solutions $\Ehat$ such that $\frob{\Bmat \Ehat - \Nmat}^2 = 0$ ($\frob{\cdot}$ denotes the Frobenius norm).
From this, one can conclude that it is generally not possible to fully recover $\Emat$ given $\Nmat$ and $\Bmat$.
However, while exact recovery of $\Emat$ may not be possible, approximations may still be possible and, if accurate (i.e., if $\Ehat \approx \Emat$), they may be sufficient for subsequent network analysis tasks (see also \cref{sec:networktasksafterrecovery}).

The classical approach to solving the underdetermined system (\ref{eqn:nodeandedgematrices}) is to find the least-norm solution~\cite{penrose_1956}.
While $\Bmat^{-1}$ in general does not exist, the pseudoinverse $\Bmat^{+}$ can be efficiently computed, using, for instance, singular value decomposition (SVD), and then used to find the minimum or least-norm solution:
$\Ehat = \Bmat^{+} \Nmat$. %
We show below (\cref{sec:understandingrecoveryerrors}) that the accuracy of information recovery using the least-norm solution depends on a combination of topological and dynamical sparsity.

In this context sparsity is also the fundamental drawback with respect to the least-norm solutions---they do not promote sparse solutions, meaning that $\Ehat$ will be a dense matrix, which is unlikely to be a good representation of $\Emat$ unless all edges are active at all times.
A period of little network activity may correspond to many zeros appearing within the time series.
Moreover, sparsity in the time series may be intermixed with sparsity in the network structure.

To address this, sparsity-promoting solution techniques, optimization problems that explicitly reward zero values in solutions, have in recent years become a pillar of statistical learning methods~\cite{tibshirani1996regression,candes2007dantzig, xu2011compressive, hastie2019statistical}.
Thus, to improve on the classical approach,
we formulate finding sparse solutions $\Ehat$ to \cref{eqn:nodeandedgematrices} as an optimization problem whose objective function consists of a least-squares error term and a regularization or penalty term that promotes zero values in discovered solutions (Supporting Information (SI) \cref{si:eqn:optimization}).
This optimization problem can be interpreted as a (multi-target) Lasso regression~\cite{tibshirani1996regression} without an intercept term and with an added constraint enforcing nonnegative regression coefficients, capturing the properties of an edge activity matrix.
We use an $L_1/L_2$ regularization to find sparse solutions for multiple problems (columns of $\Nmat$ and $\Ehat$) jointly by treating each edge as a group over time~~\cite{yuan2006model,lounici2011oracle}.
Further, this formulation is related to the similar problem of compressed sensing~\cite{lu2017binary, zhao2018sparse}.
In our case, solutions were found using coordinate descent;
see SI \cref{si:sec:findingsparsesolutions} for full details, including implementation details and Bayesian selection of the regularization hyperparameter\footnote{An implementation is available at \href{https://github.com/bagrow/recovering-information-temporal-networks}{github.com/bagrow/recovering-information-temporal-networks}.}.

To understand how well temporal network information can be recovered using the least-norm and sparse methods, we assembled a temporal network corpus consisting of five networks representing a range of different systems.
The first, `Copenhagen', used in \cref{fig:cartoon} to illustrate the information loss and recovery problem, is a social network derived from smartphone bluetooth proximity data that serve as a proxy of face-to-face interactions~\cite{sapiezynski2019interaction, stopczynski2014measuring}.
The remaining networks are 
`Hospital', another proximity-based social network derived from wearable sensors carried by healthcare workers~\cite{vanhems2013};
`Ant Colony', a physical interaction network taken from manually-annotated video footage of a \emph{Temnothorax rugatulus} colony~\cite{10.1371/journal.pone.0020298};
`Manufacturing Email', an interaction network derived from internal emails sent between employees of a mid-size manufacturing firm~\cite{michalski2011};
and `College Message', a social network derived from messages sent between members of an online community of University of California, Irving college students~\cite{panzarasa2009patterns}.
These networks cover a span of time scales, from minutes (Ant Colony) to days (Hospital), weeks (Copenhagen) and months (Manufacturing Email, College Message).
Full details for all networks, including data processing steps and network statistics, are given in SI \cref{si:sec:dataset}.

\begin{figure}[t!]
    \centering
    \includegraphics[width=\textwidth]{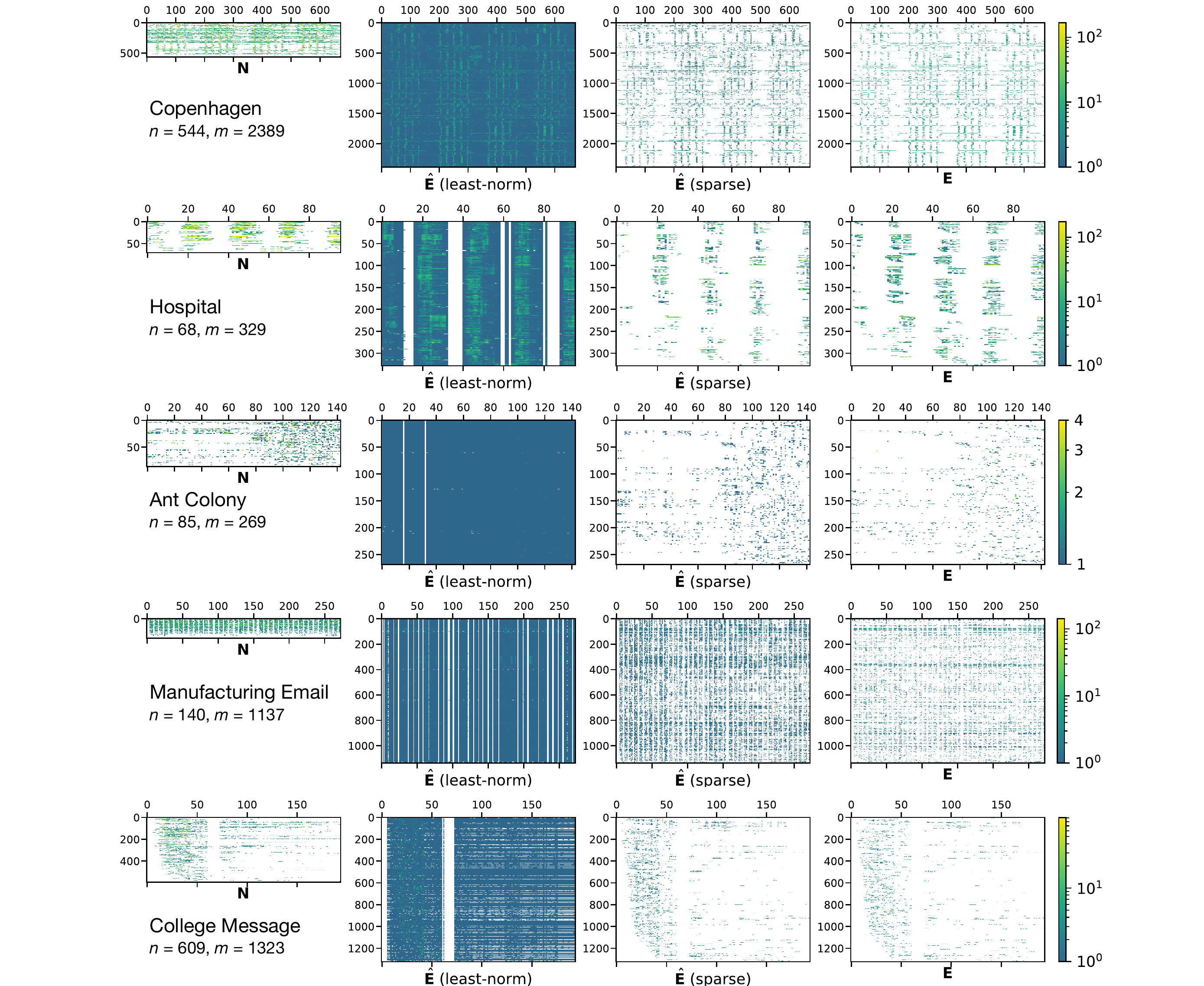}
    \caption{Good quality $\hat{\mat{E}}$ matrices are found across most networks in a representative corpus of temporal networks. 
    One notable exception is Mfg.\ Email.
    Each row corresponds to one network with the (logarithmic) color scale, indicating the magnitude of matrix elements, shared across the row. 
    Solution sparsity is illustrated with white color and is well recovered, except for Mfg.\ Email, when comparing $\hat{\mat{E}}$ (sparse) and $\mat{E}$.
        Node and edge activity matrices are drawn to scale---the small size of the $\Nmat$
compared to $\Emat$ underscores the recovery challenge.
    Least-norm solutions are dense except for time periods of zero edge activity, manifesting as vertical bands in all networks except Copenhagen; see SI for details.
    \label{fig:gridheatmaps}
    }
\end{figure}

For each network, information loss (replacing $\Emat$ with $\Nmat$) was simulated using \cref{eqn:nodeandedgematrices}.
\Cref{fig:gridheatmaps} shows heatmaps of the node activity and edge activity matrices for each network.
In the figure, rows of the node and edge matrices are drawn to scale to illustrate the extent of information loss across the corpus.
For each network, we report $n$ and $m$, the numbers of nodes and edges; their ratio $m/n$ describes the ``aspect ratio'' of the incidence matrix $\Bmat$ which reflects how much information was lost when moving from $\Emat$ to $\Nmat$ and consequently how challenging we may expect the recovery problem to be.

\Cref{fig:gridheatmaps} also shows the $\Ehat$ recovered from $\Nmat$ for both the least-norm and sparse solution methods discussed in \cref{sec:networkrecovery}.
(All matrices in a given row in \cref{fig:gridheatmaps} use the same colorbar.)
Comparing these recovered matrices to $\Emat$ shows that both approaches managed to capture the qualitative overall features of $\Emat$.
However, we see that the sparse $\Ehat$ solution is far closer in appearance to the true $\Emat$.

In particular, the sparsity pattern of $\Emat$ (zero entries in matrices are illustrated with white in \cref{fig:gridheatmaps}) is almost entirely absent in the least-norm solutions while being captured well in the sparse solutions.
Qualitatively, visual comparison of $\Ehat$ (sparse) and $\Emat$ demonstrates the potential for quite good information recovery.
Quantitatively, \cref{tab:correlations-E-Ereco} measures the recovery accuracy by reporting the correlations between matrix entries.
We report both Pearson $r$ and Spearman $\rho$ correlations to capture both linear and nonparametric associations between $\Emat$ and $\Ehat$.
Despite the challenge we would expect when attempting to recover the lost temporal information, when examining the table we see that many networks show a high correlation between the original $\Emat$ and the recovered $\Ehat$.
One exception is Manufacturing Email, where both solutions achieved correlations lower than $0.5$.
Even though it is not the largest network in the corpus, Manufacturing Email is the densest, with $m/n = 8.12$.
This relatively high density leads to increased aggregation of the edge time series' and thus we expect that undoing the subsequent information loss will be more difficult as a result.

\begin{table}%
    \centering
    \begin{tabular}{lrrrr} \toprule
                    & $r$ (ln) & $r$ (sp) & $\rho$ (ln)& $\rho$ (sp)\\ \midrule
Copenhagen          & 0.7660 & 0.5060 & 0.4512 & 0.7308 \\
Hospital            & 0.5494 & 0.5752 & 0.4067 & 0.6004 \\
Ant Colony          & 0.5698 & 0.8432 & 0.2727 & 0.6844 \\
Manufacturing Email & 0.4275 & 0.2650 & 0.3957 & 0.3879 \\
College Message     & 0.7104 & 0.8788 & 0.2508 & 0.7560 \\ \bottomrule
    \end{tabular}
    \caption{Correlations between original $\mat{E}$ and recovered $\hat{\mat{E}}$. 
    Here we report the Pearson correlation $r$ and Spearman correlation $\rho$ between elements of $\mat{E}$ and $\mat{\hat{E}}$.
    In most instances, high correlations were found, indicating successful recovery of $\Emat$, with Mfg. Email being a notable exception. 
    Often the sparse solution outperforms the (dense) least-norm solution. 
    ln: least-norm; sp: sparse.
    \label{tab:correlations-E-Ereco}
    }
\end{table}

We further explore the effects of network structure and the intersecting roles of topological and dynamical sparsity in \cref{sec:understandingrecoveryerrors}.

\section{Network tasks after recovery}
\label{sec:networktasksafterrecovery}

The recovered edge activity data will only be useful if the data are useful for subsequent tasks.
Here we examine how well the recovered data can be used on two exemplar tasks: 
estimating tie strength from temporal activity and extracting the network's multi-scale backbone~\cite{serrano2009extracting}.

Tie strengths, typically represented with edge weights $w$ accounting for the total quantity of interactions between nodes, are captured by summing $\Emat$ over time periods: $w_{ij} = \sum_t E_{ij,t}$.
When only $\Ehat$ is available, how well does $\hat{w}_{ij} \approx w_{ij}$?
We computed $\hat{w}$ using three methods. 
One, a baseline, uses a linear kernel to bypass the computation of $\Ehat$ and compute $\hat{w}$ directly from $\Nmat$:
$\hat{w}_{ij} = \mathbf{N}_{i} \mathbf{N}_{j}^{T}$, where $\mathbf{N}_i$ is the row vector corresponding to node $i$ in $\Nmat$.
The second and third methods estimate $\Ehat$ using the least-norm and sparse solution, respectively, and summing $\Ehat$ over time periods to derive $\hat{w}$.
After computing each $\hat{w}$ using $\Ehat$ we compare to $w$ derived directly from $\Emat$.

\begin{figure}[t!]
    \centering
    \includegraphics[width=0.7\textwidth]{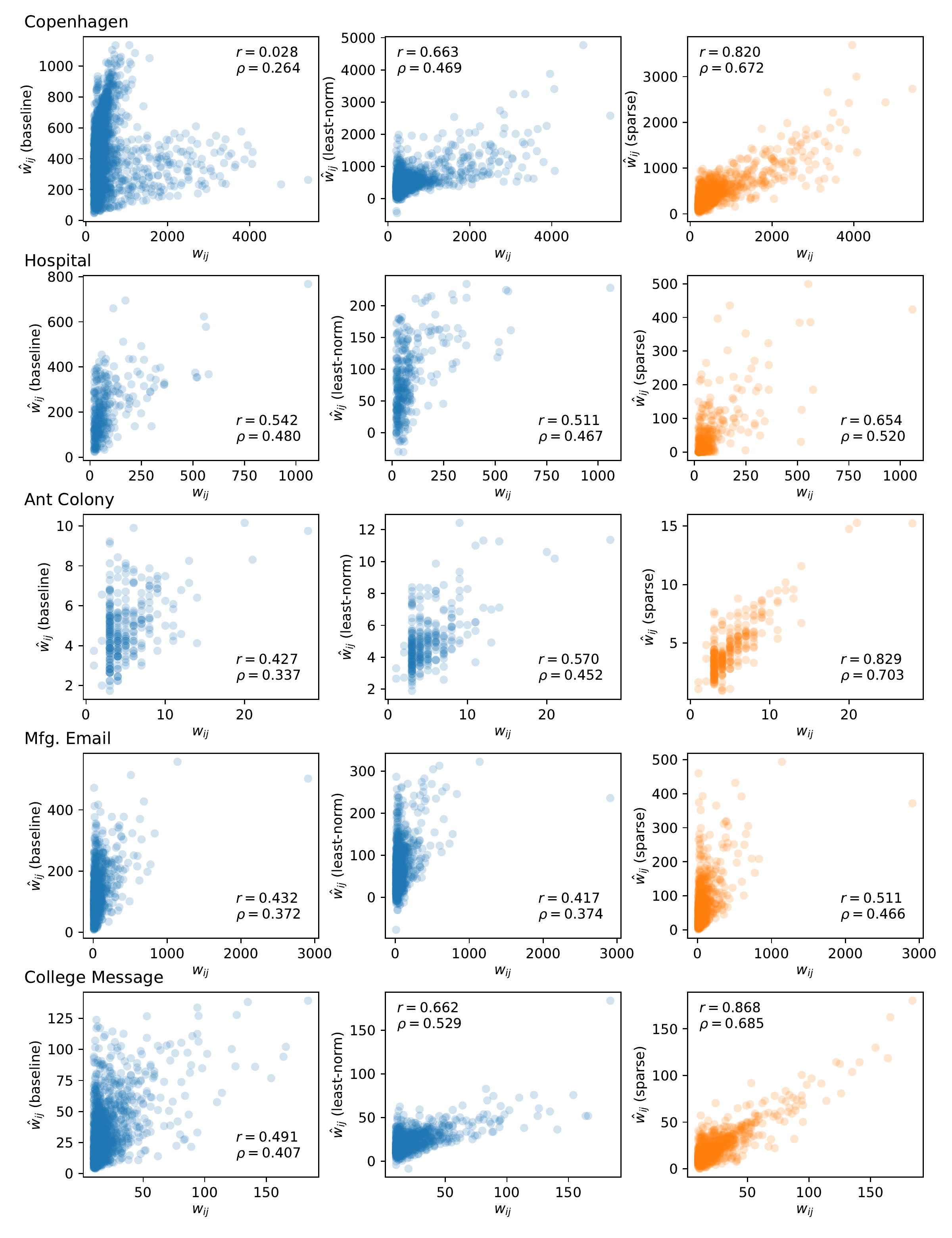}
    \caption{Quality of tie strength inference using recovered $\hat{\mat{E}}$. 
    Here we compare the edge weights $w$, computed from $\mat{E}$, with $\hat{w}$ computed directly from $\mat{N}$ (baseline) and from the least-norm and sparse $\mat{\hat{E}}$.
    In all cases, the sparse solutions (right column) best recover the edge weights.}
    \label{fig:compare-edgeweights}
\end{figure}

As shown in \cref{fig:compare-edgeweights},
good recovery of tie strength is possible across the network corpus.
In particular, the sparse solution more accurately infers the underlying tie strength than either the node-based baseline or the least-norm approach.
Three networks show correlations $r(w, \hat{w}) > 0.8$ and even the most difficult network, Manufacturing Email, shows $r>0.5$, better performance for that network than may be expected from \cref{tab:correlations-E-Ereco}.
While all networks show high recovery performance, the sparse solutions are especially accurate for the Copenhagen, Ant Colony, and College Message networks, achieving a 23.6\% (Copenhagen) to 45.4\% (Ant Colony) improvement in correlation $r$ compared to the least-norm approach.

While the high performance of the sparse solution naturally follows from its accuracy when inferring $\Emat$, in some ways, it is also surprising.
The sparsity-promoting penalty imposed on solutions induces a bias in the solution and we might expect that bias to reduce the accuracy of the aggregated quantity $\hat{w}$ as opposed to the least-norm solution where solution errors over and under $\Emat$ may cancel out during summation. 
That correlations remain high despite this further underscores the usefulness of sparse estimation.

Our second example of a post-recovery task is to extract a network's multiscale backbone~\cite{serrano2009extracting}.
The multiscale backbone represents the central or key set of edges undergirding the network's topology.
The multiscale backbone is found by examining the local distribution of edge weights~\cite{serrano2009extracting}.
Specifically, an edge ($i,j$) incident to node $i$ is flagged as belonging to the backbone if $\left(1-p_{ij}\right)^{k_i-1} < \alpha_B$, where $p_{ij} = w_{ij}/\sum_{j^\prime} w_{ij^\prime}$, $k_i$ is the degree of $i$, and $\alpha_B$ is the multiscale backbone strength parameter.
(Technically, edge $(i,j)$ is retained if this inequality is satisfied on either $(i,j)$ or $(j,i)$; see Serrano \emph{et al.} \cite{serrano2009extracting} for details.)

For each network, we compare the backbone inferred from the original data with the backbone found on the recovered data.
To do so, we interpret backbone discovery as a binary classification or partitioning of edges: each edge is flagged as either a backbone edge or non-backbone edge and we compare these classifications using false positive and true positive rates. 
However, there is an additional layer of complexity: as the multiscale backbone method depends on parameter $\alpha_B$, the ``true'' solution (i.e., the solution found on $\Emat$) will depend on $\alpha_B$ and an experimenter may not know $\alpha_B$ when working with $\Ehat$. 
We therefore distinguish between two parameters by denoting as $\alpha_B$ the value used on $\Emat$ and $\alpha_B^{\prime}$ the value used on $\Ehat$ and we consider both cases $\alpha_B = \alpha_B^{\prime}$ and $\alpha_B \neq \alpha_B^{\prime}$ when comparing results.

\begin{figure}[t!]
    \centering
    \includegraphics[width=\textwidth]{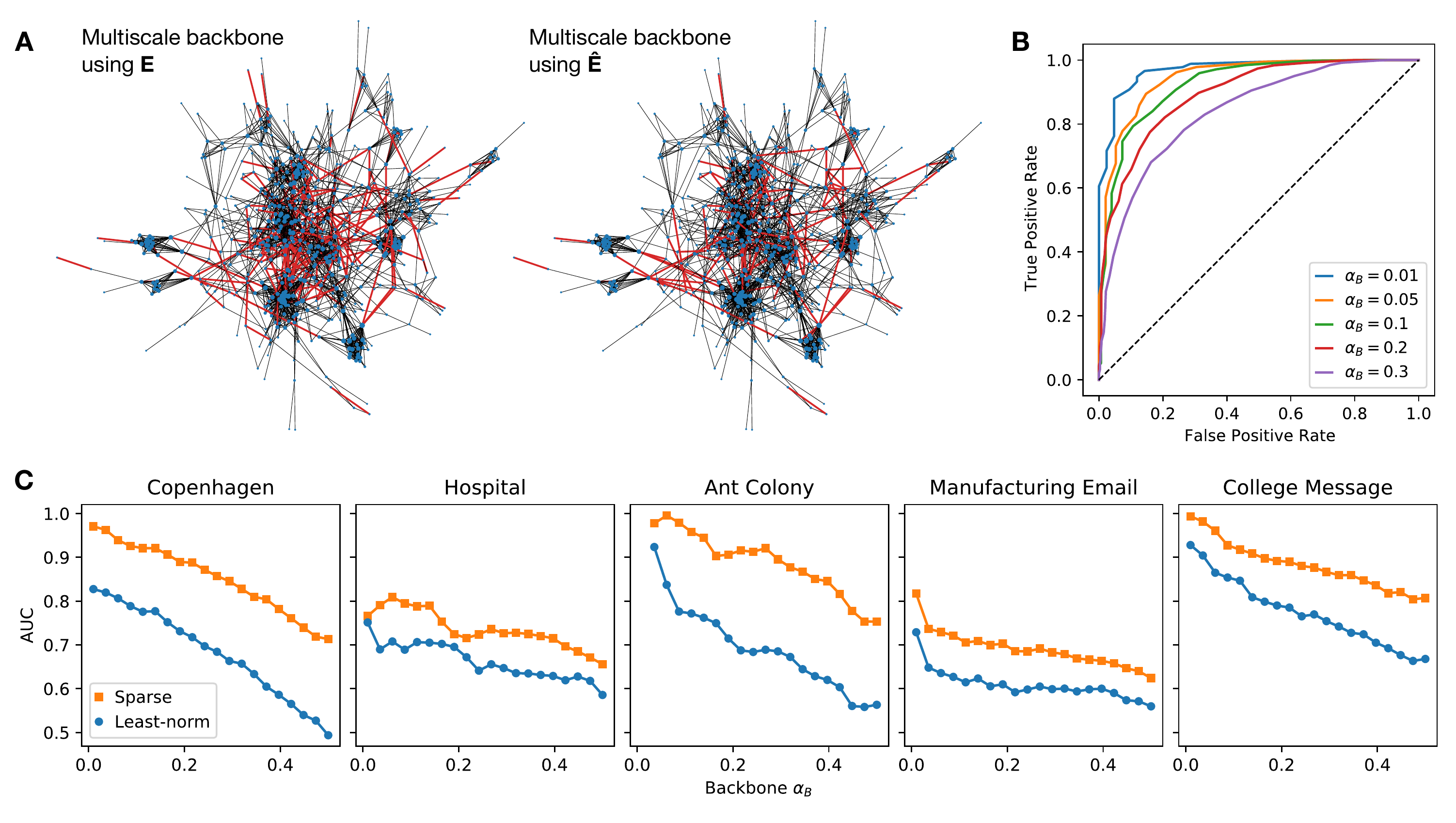}
    \caption{Extracting multiscale network backbones using recovered data.
\lett{A}
The multiscale backbones ($\alpha_B = 0.125$) using the original and recovered edge activity data for the Copenhagen social network.
Red denotes edges flagged as belonging to the backbone.
\lett{B} 
Comparing backbones inferred from the $\Ehat$ (sparse) for the Copenhagen network when $\alpha_B \neq {\alpha}_B^{\prime}$. 
Here each ROC curve corresponds to a different choice for $\alpha_B$ while the curve itself is parameterized by sweeping through values $\alpha_B^{\prime}$. 
The diagonal line denotes the null case where edges are randomly labeled as backbone or non-backbone.
ROC curves for all networks are shown in \cref{si:fig:compare-ROC-curves-edgeweight}.
\lett{C}
Expanding upon panel B to the entire corpus, we report the area under the curve (AUC) of ROC curves.
    \label{fig:compare-backbone}
    }
\end{figure}

To illustrate the extracted backbones, we begin by drawing the Copenhagen network twice in \cref{fig:compare-backbone}A, once highlighting the edges of the ``oracle'' backbone computed using $\Emat$ and again highlighting the backbone found using the sparse $\Ehat$.
While there are some visual discrepancies between the two backbones, overall they agree considerably. Moving to a quantitative assessment, in \cref{fig:compare-backbone}B we plot the Receiver Operating Characteristic (ROC) curve comparing $\Emat$-backbones and $\Ehat$-backbones on the Copenhagen network. 
The dashed line illustrates the expected classifier performance if randomly labeling edges as backbone or non-backbone; we see generally far better performance (as typified by low false positive rates and high true positive rates) regardless of $\alpha_B$ and $\alpha_B^{\prime}$.

Moving beyond the Copenhagen network, \cref{fig:compare-backbone}C shows classifier performance across the entire corpus, for both the least-norm and sparse recovered $\Ehat$.
We simplify the ROC curve by plotting the AUC (area under the ROC curve) as a function of $\alpha_B$ (individual ROC curves for all methods and networks are shown in \cref{si:sec:multiscalebackbone}, \cref{si:fig:compare-ROC-curves-edgeweight}).
Overall, we observe high AUC values, far from the AUC = $1/2$ value for a null classifier. 
Further, for any given network and value of $\alpha_B$, the sparse solution performs better at classifying backbone edges than the least-norm solution.
Performance degrades as $\alpha_B$ increases, when edges were less selectively added to the ``true'' backbone, and improves considerably at lower $\alpha_B$, when edges were more selectively added.
This trend in performance indicates that the strongest and thus most important members of the multiscale backbone can be inferred accurately even when the detailed information within $\Emat$ is lost or absent.

\section{Understanding recovery errors}
\label{sec:understandingrecoveryerrors}

We now explore the errors made when recovering $\Emat$ using the classic, least-norm and modern, sparsity-promoting solution methods applied above.
Using our network corpus, we also relate how information loss is affected by different aspects of network structure and temporal activity.

It is instructive to first consider the classic approach to solving the underdetermined system (\cref{eqn:nodeandedgematrices}).
Using the pseudoinverse $\Bmat^{+}$ to compute the least-norm solution will not be perfectly accurate:
\begin{align}
    \Ehat = \Bmat^{+} \Nmat = \Bmat^{+}\Bmat \Emat \neq \Emat
\end{align}
because $\Bmat^{+}\Bmat\neq \mat{I}$ since $\Bmat^{+}$ is a right-inverse, not a left-inverse, of $\Bmat$.
Further, since $\frob{\Bmat \Ehat - \Nmat}^2 = 0$, we cannot characterize the solution using the residual. 
Instead, we consider in theory the difference between $\Emat$ and $\Ehat$.
We show (\cref{si:sec:leastnormerrorbound}) that 
\begin{equation}
\frob{\Ehat-\Emat} \leq \sqrt{m-n} \frob{\Emat},
\label{eqn:finalminbound}
\end{equation}
where $\frob{\cdot}$ is the Frobenius norm.
In other words,
the error bound \cref{eqn:finalminbound} tells us how much our recovered edge activity matrix $\Ehat$ will be near to or far from the actual edge activity matrix $\Emat$ depends on the product of two terms, the norm of the true solution $\frob{\Emat}$ and the density of the network, specifically $\sqrt{m-n}$. 

Both terms on the right of \cref{eqn:finalminbound} relate the difficulty of the recovery problem to the sparsity of the data.
Denser networks will have larger $m$ relative to $n$, leading to a wider $\Bmat$.
Likewise, networks with denser temporal activity will have more edges active at any given time, leading to fewer zero elements in $\Emat$ and potentially larger values for non-zero elements, both of which contribute to a larger $\frob{\Emat}$.
As either form of sparsity decreases, the gap between $\Ehat$ and $\Emat$ is likely to grow, leading to larger recovery errors.

Turning from the least-norm solution, we now focus on the sparse solution method.
Sparse estimation techniques have a rich history of theoretical study~\cite{chen1994basis,candes2007dantzig,lounici2011oracle,hastie2019statistical}.
Here it is key to understand the shape of the optimization problem's loss function.
If the loss function is strongly convex, then a single minimum exists and, for optimizations such as ours, theoretical bounds can be put on the estimation error.
If, however, the loss function is not strongly convex but only convex, then there will exist directions for which the loss is flat, and the optimization solution can become arbitrarily far from the minimum.
For our optimization, at timestep $t$, i.e., column $t$ of $\Emat$ and $\Nmat$, the least-squares loss $f(\Emat_t) = \frac{1}{2n} \lVert \Nmat_t - \Bmat \Emat_t \rVert_2^{2}$ is always convex. 
It is strongly convex when the $m \times m$ Hessian matrix for this loss, $\nabla^{2}f = \Bmat^{T} \Bmat / n$, has eigenvalues bounded away from zero.
Unfortunately, a Gram matrix of the form $\Bmat^{T} \Bmat$ where $\Bmat$ is $ n \times m$ will have rank at most $\min\{n,m\}$ and so will be rank-deficient in the high-dimensional setting ($n < m$)~\cite{hastie2019statistical}.
Standard practice then is to seek weaker requirements than strong convexity for the loss function, known as \textit{restricted} strong convexity~\cite{negahban2012unified}, where strong convexity is required only for a subset 
of estimation error vectors.
For a linear model, this means that we seek to lower bound the \textit{restricted} eigenvalues (RE) of $\Bmat^{T} \Bmat$~\cite{bickel2009simultaneous}.
If zero is a eigenvalue of $\Bmat^T \Bmat$, we will be unable to guarantee a bound on the estimation error, although in practice it is still possible to find good solutions~\cite{candes2007dantzig}.

By combining results from the literature of sparse statistical learning with results from spectral graph theory, we prove (\cref{si:sec:networkstructureanderrorbounds}) that RE holds when the smallest eigenvalue of the adjacency matrix of the network's \emph{line graph} is greater than -2.
This occurs only when every connected component of the network is a tree or has one cycle only, of odd length~\cite{doob1973interrelation}.
Intuitively, this makes sense: the incidence matrix of a tree for instance will be narrower than it is wide ($m/n < 1$) and the system (\cref{eqn:nodeandedgematrices}) will not be underdetermined.
Yet this is also a strong requirement---no networks in the corpus are close to being trees. 
However, what matters for RE to hold is the network's structure at each time step, not the overall structure (see SI), and some networks, Ant Colony and College Message in particular, often meet this requirement (\cref{si:fig:mineigenvalueBmatTBmat}).
And, as in many practical situations, 
even without the theoretical guarantee, 
we have observed here successful use of recovered data in practice (\cref{fig:compare-edgeweights,fig:compare-backbone} above).

\begin{figure}[t!]
    \centering
    \includegraphics[width=\textwidth]{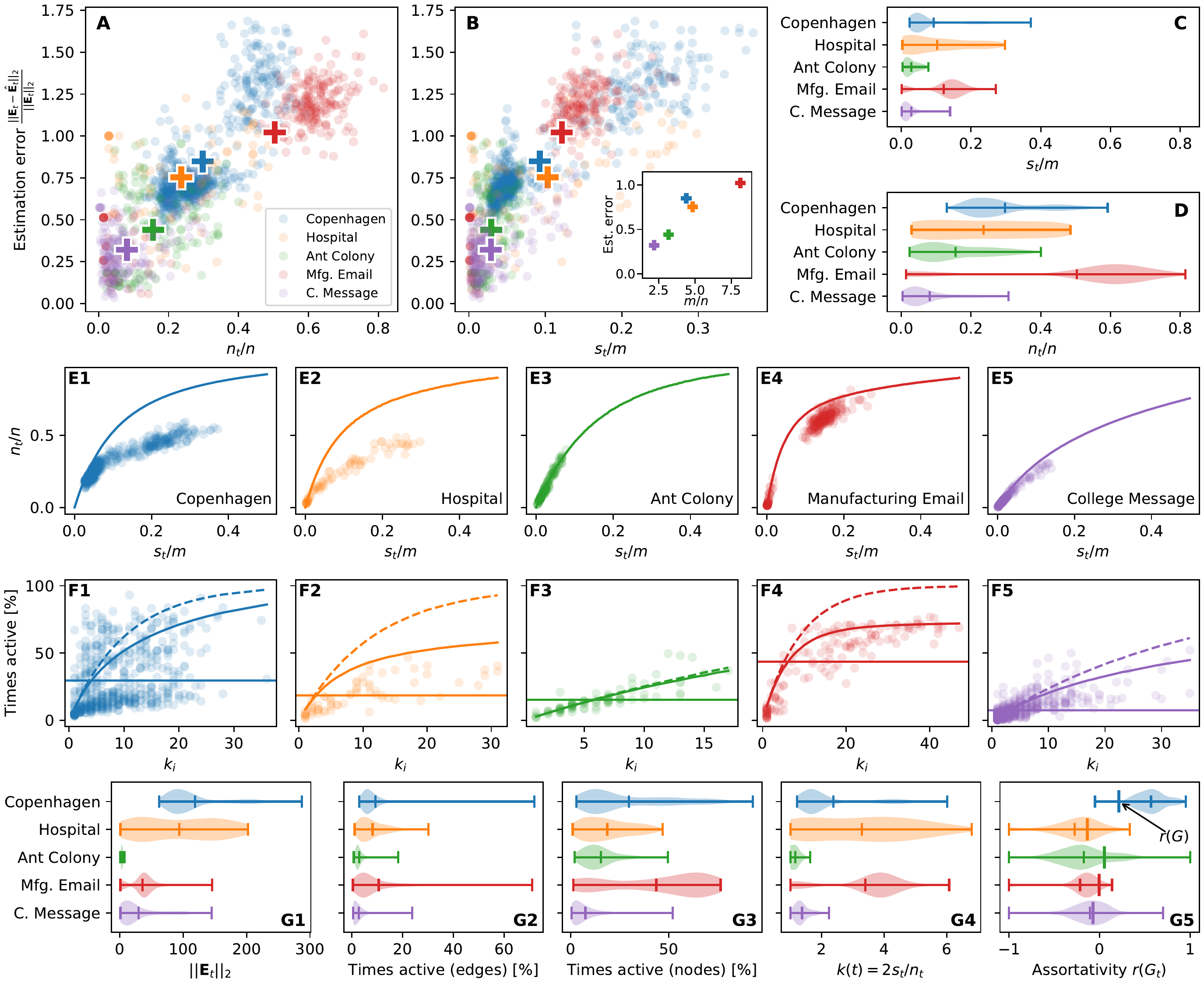}
    \caption{Topological and dynamical characteristics of recovery error.
    \lett{A} Estimation error at time $t$ depends on active node fraction $n_t/n$ and 
    \lett{B} active edge fraction $s_t/m$.
   	Crosses in A and B denote the overall average per network.
    (inset) the average estimation error is well approximated by the network density $m/n$.
    \lett{C, D} Distributions of active edges and nodes. Most networks are denser in node activity than in edge activity.
    \lett{E} Observed relation between active nodes and edges. 
    Lines denote the expected $n_t/n$ if $s_t$ distinct edges, chosen uniformly at random, are active.
    Ant Colony is well explained by this independence null, while other networks, particularly Copenhagen and Hospital deviate at high $s_t$ from the null.
    \lett{F} The proportion of times a node $i$ is active (has at least one active incident edge) as a function of its degree $k_i$. 
   	Curved lines denote expectations for null models for edge activity that preserve the overall sparsity of $E$ (dashed) and the sparsity of $E_t$ at all times $t$ (solid).
    Horizontal lines denote the mean fraction of times active.
    \lett{G} Distributions of dynamical and network features across the network corpus.
    \label{fig:error-figure}
    }
\end{figure}

Beyond the theoretical analysis, let us examine the estimation error between $\Ehat$ and $\Emat$ empirically, using our network corpus.
In \cref{fig:error-figure}A,B we plot the (relative) estimation error at time $t$ as a function of the active node $n_t/n$ and active edge fractions $s_t/m$, respectively.
Both active fractions strongly correlate with estimation errors across the network corpus.
Somewhat surprisingly, we find comparable correlations with error for both $n_t/n$ and $s_t/m$: $r = 0.831$ and $r=0.825$, respectively.
We expand further on this correlation in SI \cref{si:tab:correlations}.
Being able to use $n_t$ to judge the magnitude of errors just as well as $s_t$ is interesting because in practice the actual value of $s_t$ will not be known, but $n_t$ will be. 
A strong correlation here would allow a researcher to anticipate the magnitude of errors made during recovery, an important step to support their investigations in practice, when true solutions are unknown.
In that regard, even a simple measure of overall density, $m/n$, which is also the ``aspect ratio'' of $\Bmat$, is a good predictor for the average estimation error (\cref{fig:error-figure}B inset).

Further examining the scatter plots in \cref{fig:error-figure}A,B we see that the two sparsest networks, Ant Colony and College Message, have lower active fractions and lower recovery errors.
Conversely, the densest network, Manufacturing Email, has a very high active node fraction and high estimation errors.
Copenhagen occupies an interesting middle ground, with a bimodality in active fraction and estimation error, meaning there are time periods where the network is more strongly activated and recovery is difficult and other periods where the network's activity is sparse and recovery is less error-prone.
Examining the distributions of active fractions over time (\cref{fig:error-figure}C,D) we see that only a minority of edges are active at a given time (C) but those edges will land upon and activate a larger fraction of the nodes (D). This is most extreme for Manufacturing Email, where typically less than 20\% of active edges will give rise to 60\% active nodes.

While the distributions of node and edge activity fractions are useful to study independently, the interaction of the two also matters.
In particular, do active edges follow a pattern in their distribution over the network, or are they effectively randomly distributed?
In \cref{fig:error-figure}E we plot for each network $n_t/n$ vs.\ $s_t/m$ over $t$.
Along with these scatter plots we include a randomized null model 
where we selected the same number $s_t$ of distinct edges uniformly at random and then determined $n_t$ from how many unique nodes were incident upon that set of edges.
For small $s_t$ relative to $m$, we expect $n_t = 2 s_t$ as it will be unlikely for two edges chosen at random to fall on a common node (cf.\ graph matchings~\cite{lovasz2009matching}).
Indeed, this appears to hold for Ant Colony and to a lesser extent College Message; both are well described by the randomized null model.
In comparison, Copenhagen and Hospital, both proximity-based social networks, stay close to the null model only for the smallest values of $s_t$.
At higher values of $s_t$, both deviate significantly, indicating that edge activations cluster around a comparatively smaller set of nodes.
This clustering of active edges will make the recovery problem more challenging.
Manufacturing Email shows a similar clustering but to a lesser extent, while the high number of nodes active at the same time (cf.\ \cref{fig:error-figure}D) makes recovery a challenge.
Notice that transforming from $n_t$ and $s_t$ to $n_t/n$ and $s_t/m$ introduces a scale-dependency related to the overall density of the network:
$n_t = 2 s_t$ becomes $n_t/n = \avg{k}s_t/m$ where $\avg{k} = 2m/n$ is the average degree of the network and, further, is twice the aspect ratio of $\Bmat$. In many ways, the steepness of the curves in panel E at small $s_t$ summarizes the topological difficulty of the recovery problem.

Another view for how edge activations distribute over nodes is given in \cref{fig:error-figure}F, where we study the fraction of times a node is active as a function of its degree.
As only a single incident active edge is needed to activate a node, we anticipate that high degree hub nodes will be disproportionately activated compared to low degree nodes. 
In fact, all networks in the corpus exhibit this trend.
We also observe that the steepness of this trend correlates well with the difficulty of the recovery problem.
For comparison,
each scatter plot in \cref{fig:error-figure}F also shows two null models (curved lines) capturing the expected proportion of times active if edges are activated at random.
The first null model (dashed lines) preserves the overall number of active edges (the number of nonzero elements in $\Emat$) while the second, more strict model (solid lines) preserves the active number of edges at each time $t$ (the number of nonzero elements in each column $\Emat_t$).
Compared to the actual times active, some networks are well explained by one null model (Manufacturing Email, College Message) or even both (Ant Colony).
Other networks (Copenhagen, Hospital) are not explained by either null, indicating a non-uniformly random distribution in how active edges appear across high- and low-degree nodes.
Copenhagen is especially interesting to point out as the bimodality seen above (\cref{fig:error-figure}A,B) exhibits here as two regimes of nodes, one active far less often than expected, and another active as much or more often than expected from the null models.

Lastly, we examine a number of other dynamical and network properties in \cref{fig:error-figure}G.
\Cref{fig:error-figure}G1 shows distributions of $\lVert \Emat_t \rVert_2$, describing the (unnormalized) magnitudes of dynamical activity we can expect on edges in the network; Copenhagen, the largest network, exhibits a few periods of very high activity but otherwise exhibits a typical scale similar to Hospital.
Hospital, however, exhibits a wide, nearly uniform spread in low and high activity.
Other networks, especially the sparse Ant Colony, show less activities per timestep.
Next, supplementing the distributions of $s_t/m$ and $n_t/n$ (\cref{fig:error-figure}C,D), \cref{fig:error-figure}G2 and \cref{fig:error-figure}G3 show the distributions of times active for both nodes and links.
These follow the active fractions closely but two networks, Copenhagen and Manufacturing Email, have subsets of frequently active edges not apparent in the distributions of $s_t/m$ shown in \cref{fig:error-figure}C.
Another measure of dynamical network density, the temporal mean degree $k(t)=2s_t/n_t$ is shown in \cref{fig:error-figure}G4 (excluding times $t$ where $n_t=0$). 
Here we can see that Hospital and Manufacturing Email have similar high densities, but Manufacturing Email is consistently dense while Hospital has periods of high and low density, in line with the spread of estimation errors seen in \cref{fig:error-figure}A,B.
Finally, we report in \cref{fig:error-figure}G5 the degree assortativities~\cite{newman2003mixing} $r(G_t)$ of each network over time, where $G_t$ is the network of active nodes and edges at time $t$.
Most networks tend to be degree dissortative ($r<0$) at any given time (with Copenhagen being a notable exception), meaning that links tend to form between high and low degree nodes. 
Often the time-dependent network $G_t$ is more dissortative than the static, cumulative network $G$, with Copenhagen again being an exception.
See SI \cref{si:subsec:comparisonNetworkFeatures}, \cref{si:fig:networkpropertiesanderrors} for further comparison between network features and recovery errors.
In general, the interplay between both static and time-varying network properties (\cref{fig:error-figure}G), coupled with non-random patterns of edge activations (\cref{fig:error-figure}E,F) will influence the challenge of the recovery problem, as seen here in the per-timestep estimation errors (\cref{fig:error-figure}A,B).

\section{Discussion}
\label{sec:discussion}

We have studied the problem of information loss in temporal network data. 
We asked the commonly occurring question: when detailed edge activity data are unavailable but node activities are, can edge activities be recovered from node activities and the static network itself?
This recovery problem (generally) maps into an underdetermined linear system governed by the network structure, and we examined how to recover information loss using both classical and modern approaches to finding solutions to underdetermined systems.
Using both theoretical analyses and empirical investigations, we showed that the difficulty of information recovery is governed by a decomposition of the topological and dynamical properties of the network being investigated. 
We found that temporal networks will often have a high degree of dynamic recoverability, with surprisingly good recovery performance for multiple networks, underscoring the importance of knowing the network structure: a good picture of network structure makes it easier to determine when recovery will be accurate and when it will be inaccurate.
In particular, the density of the network structure challenges recovery, with denser networks making recovery more error-prone. 
Conversely, dynamical sparsity, where less network activity tends to occur, can make recovery more accurate.
These competing forms of density and sparsity can be exploited, and we found that sparsity-promoting methods applied to the recovery problem find good solutions, sometimes surprisingly so, for both recovery itself and for informing subsequent network analysis tasks that depend on the recovered information.

Our results here point towards a number of questions that warrant further study.
The recovery problem is an inference problem, not prediction, but it would be interesting to see how well past values of $\Nmat$ can predict future values of $\Emat$. 
As recovery errors already occur without prediction, we expect compounding errors when trying to predict $\Emat$ unless the system under study happens to exhibit stationarity.
Likewise, what are the effects of measurement error?
In other words, how much more difficult is recovery when $\Bmat$ is not known perfectly?
This question should be addressed and is an important direction for future work.
Errors-in-variables methods are the natural starting point. 
Although this problem may be especially challenging as sparse estimators are known to be unstable under uncertainty, the use of specialized estimators may be fruitful~\cite{rosenbaum2010sparse}. 
And local recovery, where only subgraph dynamics are considered, are worth investigating, as local problems are often sufficient around network regions of interest and those subgraphs may be more tractable to analyze than the global network, both from a data collection perspective and from the perspective of the recovery problem.

This work underscores how important it is, across a variety of problem domains, to know the underlying network structure, as the incidence matrix feeds into the solution methods we consider.
But this raises an important question: %
what are the scientific consequences and broader impacts, both positive and negative, that may arise as a result of gathering static network data and then recovering, with perhaps unexpected accuracy, temporality?

We discuss two scenarios.
First, consider the example of an application (app) maker distributing their application on smartphones.
Smartphones contain many sensors and collect much sensitive data~\cite{sipior2014privacy,kroger2019privacy}, such as location information~\cite{de2013unique}, but also have privacy safeguards in place~\cite{7113562}.
While a potentially rich source of social network and human behavioral data, outside their own application, the application maker will not have access to sensitive information such as communication records.
Now suppose that app maker requests access to user address books. 
These address books constitute local, egocentric snapshots of the users' static social network, and are especially informative after disambiguation and record linkage across datasets~\cite{shu2017linkageAcross}.
When combined with sensor data or other records, therefore, these static networks snapshots can be used to infer not just the intrinsic activities of users (the node activities) but potentially even the activities of specific social ties (the edge activities).
While inferences along these lines will certainly require significant data, and are especially interesting in certain use cases, such as inferring the illicit communications of criminals,
this scenario underscores the privacy concerns here, as it means an entity such as a social media platform provider can begin to infer more information than expected about individuals by distributing applications and cross-referencing multiple datasets.
For the second scenario, we consider the implications for network neuroscience~\cite{bassett2017network}.
Imaging studies reveal spatiotemporal dynamics, generating
time series that act as $\Nmat$ for an aggregated network where nodes represent brain regions-of-interest~\cite{van2012human}. 
More useful would be to capture the edge dynamics governing when different ROIs ``talk'' to one another; $\Emat$ is not accessible.
However, our results here imply that, as network structure is revealed, for instance with tract- or fiber-tracing methods such as diffusion spectrum imaging~\cite{wedeen2005mapping,menzel2011accelerated} or fluorescence microscopy~\cite{livet2007transgenic}, then $\Bmat$ can be inferred and combined with $\Nmat$ to approximate the missing $\Emat$.
Accurate inference of these hidden dynamics can inform studies that move beyond the structural connectome and explore the brain's dynamical ``dynome''~\cite{kopell2014beyond}.

\section*{Acknowledgments}

We are grateful to 
M.~Almassalkhi
and
H.~Ossareh
for useful discussions. 
JPB acknowledges support by Google Open Source under the Open-Source Complex Ecosystems And Networks (OCEAN) project, U.S. Department of Energy's Advanced Research Projects Agency - Energy (ARPA-E) award DE-AR0000694,
NASA under grant 80NSSC20M0213,
and by the Broad Agency Announcement Program and Cold Regions Research and Engineering Laboratory (ERDC-CRREL) under Contract No.~W913E521C0003.
SL thanks the Villum Foundation (Nation Scale Social Networks) for support.
{\singlespacing

}



\section*{Supplementary material}

# Supporting Information

## Recovering lost and absent information in temporal networks

July 22, 2021

James P. Bagrow and Sune Lehmann

## Contents

## List of Figures

## List of Tables

## S1 Dataset descriptions

For this study we utilize five temporal network datasets taken across a variety of scientific domains:

**Copenhagen** A temporal physical proximity network tracking the smartphones of individuals participating in the Copenhagen Networks study [1, 2]. Edges were recorded as timestamped bluetooth interaction between participants. Interactions with devices outside the experiment were ignored. To process these data, edges with at least 200 interactions were retained and temporal interactions were binned from 5 minutes and then aggregated into 1h bins. Edge directionality was not used. After processing, the network consisted of $n = 544$ nodes, $m = 2389$ edges, and $T = 672$ time periods. The original data are available at https://doi.org/10.6084/m9.figshare.7267433.

**Hospital** A temporal physical proximity network tracking patient-to-patient, patient-to-worker, and worker-to-worker contacts within a hospital ward in Lyon, France [3]. Health care workers and patients wore wearable sensors (RFID) that recorded close proximity and undirected edges were added to the temporal network when an close interaction was recorded (i.e., two wearables exchanged a packet) during a 20s window. The study covered the time period Monday, Dec. 6, 2010 1pm to Friday, Dec. 10 2pm. To process these data, edges with at least 20 interactions were retained and temporal interactions were aggregated into 1h bins. After processing, the network consisted of $n = 68$ nodes, $m = 329$ edges, and $T = 96$ time periods. The original data are available at http://www.sociopatterns.org/datasets/hospital-ward-dynamic-contact-network/.

**Ant Colony** A temporal physical proximity network tracking interactions between ants (*Temnothorax rugatulus*) [4]. Edges were recorded between ants when one ant's antenna made contact with another ant's body. The colony was housed in a transparent nest and filmed; individual ants were distinguished by manual marks made with colored paint. Here we focused on Colony 1-1, the first filming period for colony 1. Filming sessions of all colonies covered periods of approximately 1800s. Data were provided as PDF table files which were extracted to plain text csv files using the Camelot Python package. To process these data, edges with at least 3 interactions were retained and temporal interactions were aggregated into 10s bins. Edge directionality was not used. After processing, the network consisted of $n = 85$ nodes, $m = 269$ edges, and $T = 143$ time periods. The original data are available at https://doi.org/10.1371/journal.pone.0020298.s008.

**Manufacturing Email** A temporal social network tracking internal emails sent between employees of a mid-size manufacturing firm [5]. Edges were recorded when emails were sent between email addresses; emails with multiple recipients (including CC and BCC recipients) were recorded as multiple simultaneous edges. The data covered the period Jan. 1, 2010 to Sep. 30, 2010. To process these data, edges with at least 10 interactions were retained and temporal interactions were aggregated into 1 day bins. Edge directionality was not used. After processing, the network consisted of $n = 140$ nodes, $m = 1137$ edges, and $T = 273$ time periods. The original data are available at https://doi.org/10.7910/DVN/6Z3CGX.

**College Message** An temporal social network originating from an online community of college students attending University of California, Irvine [6]. Timestamped links were recorded when messages were sent between students during the period April to October, 2004. To process these data, edges with at least 10 interactions were retained and temporal interactions were aggregated into 1 day bins. Edge directionality was not used. After processing, the network consisted of $n = 609$ nodes, $m = 1323$ edges, and $T = 194$ time periods. The original data are available at https://toreopsahl.com/datasets/#online_social_network.

## S1.1 Time series of network sparsity

Figure S1 shows the sparsity patterns of the temporal networks over time. The five representative networks all display different patterns or time scales. For instance, the interaction networks (Copenhagen and Hospital) display a daily periodicity, with Copenhagen covering four weeks compared to Hospital's four days. In contrast, Ant Colony covers a short time period of 30 minutes while Manufacturing Email and College Message both cover several months of time. College Message also exhibits a non-stationary signature not present in the periodicity of Manufacturing Email with a burst of activity early on followed by a long period of low activity. All networks but Copenhagen exhibit a few time periods of zero activity (highlighted in Fig. S1 with red vertical bands).

## S2 Finding sparse solutions

The system $\mathbf{BE} = \mathbf{N}$ is underdetermined, but sparse solutions $\hat{\mathbf{E}}$ for the unknown $\mathbf{E}$ can be found by solving the following optimization problem (Eq. (S1)):

$$\min_{\mathbf{E}} \quad \frac{1}{2n} \|\mathbf{N} - \mathbf{BE}\|_F^2 + \alpha \|\mathbf{E}\|_{21} \tag{S1a}$$

$$\text{subject to} \quad E_{ij} \geq 0, i = 1, \ldots, m, j = 1, \ldots, T, \tag{S1b}$$

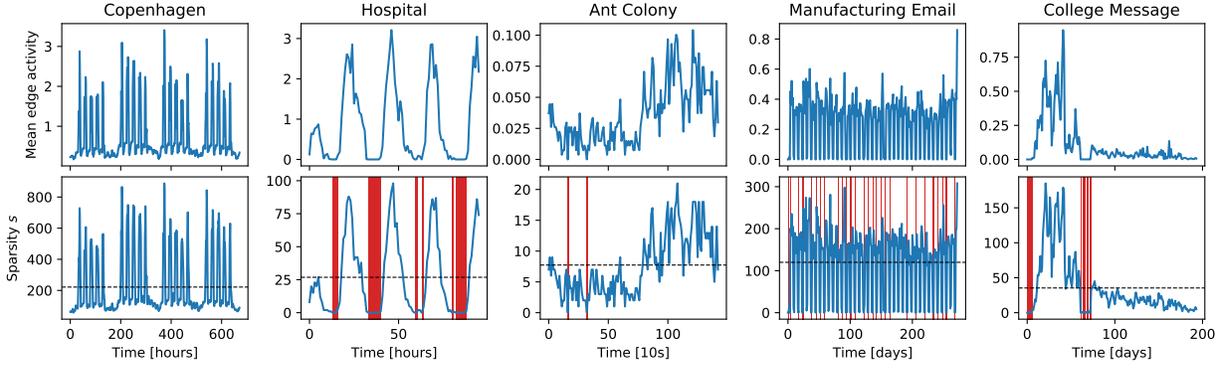

Figure S1: Sparsity over time across network datasets. (Top row) Mean edge activity $\langle \mathbf{E}_t \rangle$ over time, where $\mathbf{E}_t$ is the $t$-th column of the edge activity matrix $\mathbf{E}$. (Bottom row) Sparsity $s_t$, the number of active edges at time period $t$. Vertical red lines highlight $t$ where $s_t = 0$; cf. Fig. 2, second column.

where $\|\cdot\|_F$ is the Frobenius norm, $\|\mathbf{E}\|_{21} = \sum_i \sqrt{\sum_j E_{ij}^2}$ is a mixed $L_1/L_2$ regularization term and $\alpha$ serves as a hyperparameter balancing sparsity of $\hat{\mathbf{E}}$ and solution accuracy (via the least-squares loss $\|\mathbf{N} - \mathbf{B}\mathbf{E}\|_F$). The constraint $\mathbf{E} > 0$ enforces that edge activity data are nonnegative. (The problem Eq. (S1) can be equivalently expressed with the regularization term as a constraint.)

Equation (S1) can be interpreted as a (multi-target) Lasso regression without an intercept term and with the added constraint enforcing nonnegative regression coefficients. The $L_1/L_2$ regularization term is used to find sparse solutions for multiple regression problems (columns of $\mathbf{N}$ and $\hat{\mathbf{E}}$) jointly by treating each edge as a group over time [7, 8].

Solutions were found in Python version 3.8.8 using the Lasso implementation of Sci-Kit Learn version 0.24.1 [9]. Coordinate descent was used with a precomputed Gram matrix and random coefficient selection to speed convergence. Bayes Information Criterion was used to select $\alpha$ (Sec. S2.1). All other parameters were left to their v0.24.1 defaults.

## S2.1 Hyperparameter selection

Solving Eq. (S1) requires first determining a value of $\alpha$, the regularization hyperparameter. Different values of $\alpha$ can lead to very different solutions, so care must be taken to find the most appropriate value of $\alpha$ possible. Generally, in a machine learning or statistical learning context, it is best to $\alpha$ using additional held-out data, to help ensure your fitted model generalizes to new data. However, in our problem, we are performing inference, not learning, and while it is straightforward to use cross-validation to select $\alpha$ and fit the model with different data subsets, we can safely use the same data for both selecting $\alpha$ and estimating $\hat{\mathbf{E}}$.

Information criteria are an effective way to estimate hyperparameters [10, 11]. These balance the complexity of the model with its accuracy. For regularized regression, model complexity maps to the number of nonzero regression coefficients, which is controlled by the value of $\alpha$. This allows us to select the value of $\alpha$ by minimizing the BIC: Here we use Bayes Information Criterion (BIC):

$$\mathrm{BIC} = \frac{nT \left\langle \left( \mathbf{N} - \mathbf{B}\hat{\mathbf{E}} \right)^2 \right\rangle}{\sigma^2} + \log(nT)\widehat{df} \tag{S2}$$

where $\left\langle (\mathbf{N} - \mathbf{B}\hat{\mathbf{E}})^2 \right\rangle$ is the mean squared error, $\sigma^2 = \mathrm{Var}(\mathbf{N})$ is the variance of elements of $\mathbf{N}$, and $\widehat{df}$ is the number of nonzero coefficients, an unbiased estimator of the Lasso degrees of freedom [12]. Equation (S2) is a function of $\alpha$ for a given temporal network dataset because the data are fixed and $\hat{\mathbf{E}}$ depends on $\alpha$.

Figure S2 plots the BIC as a function of $\alpha$ for each network under study. Table S1 displays the final selected $\alpha$ for each network. Solving Eq. (S1) many times can be costly; fortunately, the path of all solutions can be computed

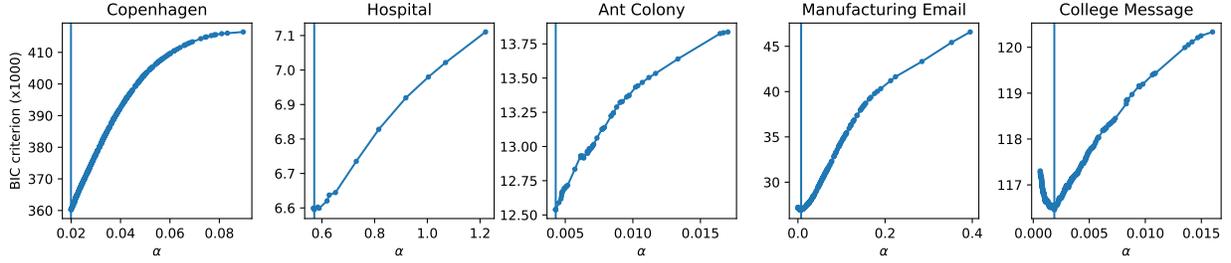

Figure S2: Hyperparameter selection using Bayes Information Criterion (BIC). For each network we plot the BIC curve as a function of the $\alpha$ regularization parameter. The smallest value and our final choice for the regularization parameter is highlighted with a vertical line. BIC was computed along each piecewise-linear step of the LARS solution path and the $\alpha$ minimizing BIC was selected. Note that the non-negativity constraint in Eq. (S1) means that the solution path does not converge to least-squares as $\alpha \to 0$. Here we report only those $\alpha > 0$ that were reached by the step-wise Lars path as only those coefficients are in congruence with the coordinate descent Lasso.

efficiently using the LARS method [13]. Sci-kit Learn's current LARS implementation does not support multi-target regression problems. To counter this, we reshaped the problem into a single regression by stacking the $T$ columns of $\mathbf{N}$ into a single vector, replicating $\mathbf{B}$ $T$ times to match, and computing the LARS solution path using an $L_1$ penalty. Then BIC was computed along the LARS solution path to select $\alpha$, which was then substituted back into Eq. (S1) and optimized to yield our final solution $\hat{\mathbf{E}}$.

Table S1: Regularization hyperparameters selected with BIC.

|  | $\alpha$ (BIC) |
| --- | --- |
| Copenhagen | 0.019958537790598734 |
| Hospital | 0.5710784313725501 |
| Ant Colony | 0.004292537772684784 |
| Manufacturing Email | 0.007326007326005932 |
| College Message | 0.0018803877107223018 |

## S3  Multi-scale backbones

Supplementing main text Fig. 4, Fig. S3 shows the least-norm solution and sparse solution ROC curves for all networks in the corpus. The main text figure only showed the sparse solution ROC curve for one network (Copenhagen) and then used plots of the AUC to summarize performance across networks and both solutions.

## S4  Least-norm error bound

Here we derive the bound on estimation error for the classic, least-norm approach to solving the underdetermined system $\mathbf{BE} = \mathbf{N}$ discussed in the main text.

While $\mathbf{B}^{-1}$ in general does not exist, the pseudoinverse $\mathbf{B}^+$ can be efficiently computed with SVD and used to find the minimum or least-norm solution [14]:

$$\hat{\mathbf{E}} = \mathbf{B}^+\mathbf{N} = \mathbf{B}^+\mathbf{B}\mathbf{E} \neq \mathbf{E} \tag{S3}$$

where the last line follows from $\mathbf{B}^+\mathbf{B} \neq \mathbf{I}$ since $\mathbf{B}^+$ is a right-inverse, not a left-inverse, of $\mathbf{B}$. Since $\left\|\mathbf{B}\hat{\mathbf{E}} - \mathbf{N}\right\|_F^2 = 0$

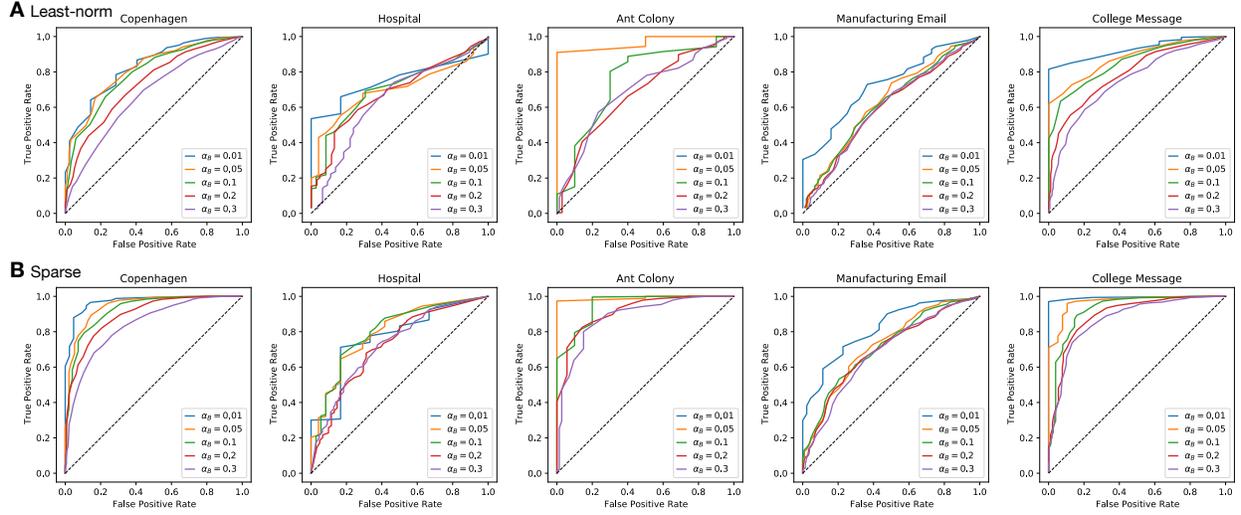

Figure S3: Receiver operating characteristic (ROC) curves comparing multiscale backbones for full corpus. This supporting plot supplements main text Fig. 4B which shows the ROC curve only for the Copenhagen network. (**A**) Backbones found with inferred edge weights recovered from the least-norm solution. (**B**) Backbones found with inferred edge weights recovered from the sparse solution. The diagonal dashed line corresponds to a classifier that is randomly guessing whether or not an edge is within the backbone; better-performing classifiers have ROC curves that push further into the upper-left corner of the ROC plot.

(assuming $\hat{\mathbf{E}}$ exists), we cannot characterize the solution using the residual. Instead, at least theoretically, we consider the estimation error, the difference between $\mathbf{E}$ and $\hat{\mathbf{E}}$:

$$\hat{\mathbf{E}} - \mathbf{E} = \mathbf{B}^+\mathbf{B}\mathbf{E} - \mathbf{E} = \left(\mathbf{B}^+\mathbf{B} - \mathbf{I}\right)\mathbf{E} \tag{S4}$$

$$\left\|\hat{\mathbf{E}} - \mathbf{E}\right\|_F = \left\|\left(\mathbf{B}^+\mathbf{B} - \mathbf{I}\right)\mathbf{E}\right\|_F \tag{S5}$$

$$\leq \left\|\left(\mathbf{B}^+\mathbf{B} - \mathbf{I}\right)\right\|_F \left\|\mathbf{E}\right\|_F \tag{S6}$$

$$\leq \sqrt{m-n}\,\left\|\mathbf{E}\right\|_F \tag{S7}$$

where $\|\cdot\|_F$ is the Frobenius norm, Eq. (S6) follows from the Cauchy-Schwarz inequality, and we now derive $\|(\mathbf{B}^+\mathbf{B} - \mathbf{I})\|_F \leq \sqrt{m-n}$.

Recall the (unoriented) incidence matrix $\mathbf{B}$ is an $n \times m$ matrix with right-inverse $\mathbf{B}^+$ such that $\mathbf{B}^+\mathbf{B}$ is $m \times m$. (We assume $\mathbf{B}$ is full rank; otherwise, either the problem is not underdetermined and can be solved or the network is disconnected and we can find the solution individually for each connected component.) The Frobenius norm of a matrix $\mathbf{A}$ is $\|\mathbf{A}\|_F^2 = \mathrm{Tr}\left(\mathbf{A}\mathbf{A}^*\right)$. Therefore

$$\begin{aligned}
\left\|\mathbf{B}^+\mathbf{B} - \mathbf{I}\right\|_F^2 &= \mathrm{Tr}\left[\left(\mathbf{B}^+\mathbf{B} - \mathbf{I}\right)\left(\mathbf{B}^+\mathbf{B} - \mathbf{I}\right)^*\right] \\
&= \mathrm{Tr}\left[\left(\mathbf{B}^+\mathbf{B} - \mathbf{I}\right)\left(\left(\mathbf{B}^+\mathbf{B}\right)^* - \mathbf{I}\right)\right] && \text{(since } \mathbf{I}^* = \mathbf{I}\text{)} \\
&= \mathrm{Tr}\left[\left(\mathbf{B}^+\mathbf{B} - \mathbf{I}\right)\left(\mathbf{B}^+\mathbf{B} - \mathbf{I}\right)\right] && (\mathbf{B}^+\mathbf{B} \text{ is Hermitian so is } \mathbf{B}^+\mathbf{B} - \mathbf{I}) \\
&= \mathrm{Tr}\left[\mathbf{B}^+\mathbf{B} - \mathbf{B}^+\mathbf{B} - \mathbf{B}^+\mathbf{B} + \mathbf{I}^2\right] && (\mathbf{B}\mathbf{B}^+ = \mathbf{I} \text{ since } \mathbf{B}^+ \text{ is a right-inverse of } \mathbf{B}) \\
&= \mathrm{Tr}\left[\mathbf{I} - \mathbf{B}^+\mathbf{B}\right] \\
&= \mathrm{Tr}\,\mathbf{I} - \mathrm{Tr}\left(\mathbf{B}^+\mathbf{B}\right) \\
&\leq m - n. && (\mathbf{B}^+\mathbf{B} \text{ is idempotent so } \mathrm{Tr}\left(\mathbf{B}^+\mathbf{B}\right) = \mathrm{rank}\left(\mathbf{B}^+\mathbf{B}\right) = \mathrm{rank}(\mathbf{B}) \leq n, \text{ since } \mathrm{rank}(\mathbf{B}) \leq n)
\end{aligned}$$

*Remark.* In general, we can expect $\mathrm{rank}(\mathbf{B}) = n$ for most networks; however, this is not guaranteed. Indeed, a connected tree will have $\mathrm{rank}(\mathbf{B}) = n - 1$. Likewise, the rank of the incidence matrix is also less than $n$ for disconnected networks.

# S5 Network structure and error bounds for sparse solutions

Here we exploit theory developed for high-dimensional ($m \gg n$) statistical learning and optimization problems to understand when the estimation error $\|\mathbf{E}_t^* - \hat{\mathbf{E}}_t\|_2^2$ can be bounded. Problems in this area include basis pursuit, compressed sensing, and penalized regression [15, 16, 17]. We connect this theory with results from spectral graph theory to determine what network structure is necessary for the error of the recovery problem to be bounded.

To understand the performance of the optimization problem we need to investigate its loss function. If the loss function is strongly convex, then a single minima exists and, for our optimization problem, theoretical bounds can be put on the estimation error. If, however, the loss function is not strongly convex but only convex, then there will exist directions for which the loss is flat, and the optimization solution can become arbitrarily far away, preventing convergence.

For our optimization (Eq. (S1)), consider, at timestep $t$, i.e., column $t$ of $\mathbf{E}$ and $\mathbf{N}$, the least-squares loss $f(\mathbf{E}_t) = \frac{1}{2n}\|\mathbf{N}_t - \mathbf{B}\mathbf{E}_t\|_2^2$ is always convex. It is strongly convex when the $m \times m$ Hessian matrix for this loss, $\nabla^2 f = \mathbf{B}^T\mathbf{B}/n$, has eigenvalues bounded away from zero. Unfortunately, a Gram matrix of the form $\mathbf{B}^T\mathbf{B}$ where $\mathbf{B}$ is $n \times m$ will have rank at most $\min\{n, m\}$ and so will be rank-deficient in the high-dimensional setting ($n < m$) [18].

This forces us to seek weaker requirements than strong convexity for the loss function. Instead, we consider *restricted* strong convexity, where strong convexity is required only for a subset $C \subset \mathbb{R}^m$ of estimation error vectors $\nu \equiv \mathbf{E}_t^* - \mathbf{E}_t \in C$. A loss function $f(\mathbf{E}_t)$ satisfies restricted strong convexity at $\mathbf{E}_t^*$ if

$$\frac{\nu^T \nabla^2 f(\mathbf{E}_t)\nu}{\|\nu\|_2^2} \geq \gamma \text{ for all nonzero } \nu \in C, \tag{S8}$$

and for all $\mathbf{E}_t \in \mathbb{R}^m$ in the neighborhood of $\mathbf{E}_t^*$.

Specializing this to a linear model means that we seek to lower bound the *restricted* eigenvalues of $\mathbf{B}^T\mathbf{B}$:

**Definition S5.1** (Restricted eigenvalue (RE) condition for linear regression).

$$\frac{\frac{1}{n}\nu^T \mathbf{B}^T\mathbf{B}\nu}{\|\nu\|_2^2} \geq \gamma \text{ for all nonzero } \nu \in C. \tag{S9}$$

Consider the regularized form of the Lasso[1]. The restricted set $C$ for the regularized Lasso is defined based on the sparsity of the parameter $\mathbf{E}_t^*$. Suppose the "active set" of edges at time $t$ is $S_t(\mathbf{E}_t)$, with $|S_t| = s_t$. For the Lasso error $\nu_t = \mathbf{E}_t^* - \hat{\mathbf{E}}_t$, let $\nu_{t,S}$ be the subvector of $\nu_t$ with elements corresponding to index set $S_t$, the nonzero elements of $\mathbf{E}_t$. Define the complement likewise: $\nu_{t,S^C}$. Then, the set of error vectors for regularized Lasso, depending on the regularization parameter [18], will satisfy a cone constraint:

$$C(S) \equiv \left\{\nu \in \mathbb{R}^m \mid \|\nu_{S^C}\|_1 \leq 3 \|\nu_S\|_1\right\}. \tag{S10}$$

Moving on to the network structure, $\mathbf{B}$ is the $n \times m$ unoriented incidence matrix of the graph $G$ and this means our regression problem is related to the *line graph* of $G$:

**Definition S5.2** (Line graph). The line graph $L(G)$ of a graph $G$ is the graph where each node in $L(G)$ corresponds to an edge in $G$, and an edge exists between two nodes in $L(G)$ when the corresponding edges in $G$ share an endpoint. The adjacency matrix $\mathbf{A}(L(G))$ is of size $m \times m$ for $G$ with $m$ edges.

Specifically, the regression problem is related to the line graph by

$$\mathbf{B}^T\mathbf{B} = 2\mathbf{I} + \mathbf{A}(L(G)). \tag{S11}$$

Each eigenvalue of $\mathbf{B}^T\mathbf{B}$ is $2 + \lambda_j$, where $\lambda_j$ is an eigenvalue of $\mathbf{A}(L(G))$. Therefore, we seek to understand the

---

[1]For simplicity, here we consider the per-timestep Lasso, which is closely aligned with the group Lasso formulation we used. When analyzing each time period $t$ separately (considering column $\mathbf{E}_t$ of $\mathbf{E}$), the optimization problem Eq. (S1) reduces to regularized Lasso (barring the non-negativity constraints).

eigenvalues of $\mathbf{A}(L(G))$ for a given graph $G$, specifically when the minimum eigenvalue $\lambda_{\min} \equiv \min_j \lambda_j > -2$. Taking $G = G_t$, the graph of active edges at time $t$, we can then use $\lambda_{\min}$ to determine if RE holds at time $t$.

To proceed, first we state two theorems due to Doob [19]:

**Theorem S5.1.** *The smallest eigenvalue $\lambda_{\min}$ of the adjacency matrix of the line graph of $G$ is*

$$\lambda_{\min}(\mathbf{A}(L(G))) = -2 \tag{S12}$$

*unless every connected component of $G$ is a tree or has one cycle only, of odd length.*

**Theorem S5.2.** *If $G$ is a graph of diameter $d$, then*

$$-2 \leq \lambda_{\min}(\mathbf{A}(L(G))) \leq -2\cos\left(\frac{\pi}{d+1}\right) \tag{S13}$$

*Further, these bounds are the best possible.*

If there exists an eigenvector $\mathbf{x}$ of $\mathbf{B}^T\mathbf{B}$ with eigenvalue zero, and $\mathbf{x} \in C$, then the estimation error will not be bounded. In fact, we need only to consider the submatrix $\mathbf{B}_t$ indexed by the active edge set $S_t$, i.e., the incidence matrix of $G_t$:

**Theorem S5.3.** *Every eigenvector of $\mathbf{B}_t^T\mathbf{B}_t$ has a corresponding vector in $C(S_t)$.*

*Proof.* Let $\mathbf{x} \in \mathbb{R}^{s_t}$ be an eigenvector of $\mathbf{B}_t^T\mathbf{B}_t$. Using $S_t \subset [1..m]$, $|S_t| = s_t$, as an index set, define $\mathbf{x}' \in \mathbb{R}^m$ to be the vector with elements

$$x'_j = \begin{cases} x_{\sigma(j)} & j \in S_t, \\ 0 & \text{otherwise}, \end{cases} \tag{S14}$$

where $\sigma : S_t \to [1..s_t]$ maps vector indices of $\mathbb{R}^m$ to those of $\mathbb{R}^{s_t}$, as subsetted by $S_t$. In other words, $\mathbf{x}'$ is the vector with active set $\mathbf{x}'_{S_t} = \mathbf{x}$ and the complement $\mathbf{x}'_{S_t^C} = \mathbf{0}$. By the construction of $S_t$, $\|\mathbf{x}'_{S_t}\|_1 = \|\mathbf{x}\|_1 > 0$, and so

$$\left\|\mathbf{x}'_{S_t^C}\right\|_1 < \delta \left\|\mathbf{x}'_{S_t}\right\|_1$$

for any $\delta > 0$. Therefore, the cone constraint of $C(S_t)$ is satisfied (with $\delta = 3$) and $\mathbf{x}' \in C(S_t)$. □

In summary, Theorem S5.3 tells us that we do not need to consider the eigenvector corresponding to $\lambda_{\min}$; it suffices only to determine whether $\lambda_{\min}(\mathbf{A}(L(G_t))) > -2$ per Thm. S5.1 and if so then the smallest eigenvalue of $\mathbf{B}^T\mathbf{B} > 0$, RE holds at time $t$, and the estimation error $\|\mathbf{E}_t^* - \hat{\mathbf{E}}_t\|_2$ will be bounded.

*Remark.* While a lower bound on $\lambda_{\min}$ tighter than $\lambda_{\min} > -2$ would be useful when Thm. S5.1 holds, Thm. S5.2 can be used to specify an upper bound on $\lambda_{\min}$. Of course, given an observed $G_t$, one can compute the smallest eigenvalue of $\mathbf{B}^T\mathbf{B}$ directly to determine $\gamma$ and apply an appropriate bound. Bounds on estimation (and prediction) error in high-dimension settings are the subject of considerable research in many contexts [20, 17, 21, 22, 8, 23]. See [18, 24] for recent summaries.

*Remark.* If $G_t$ is disconnected, $L(G_t)$ will also be disconnected and $\mathbf{A}(L(G_t))$ will be reducible. Because $L(G_t)$ is undirected, $\mathbf{A}(L(G_t))$ can further be put into block diagonal form when $G_t$ is disconnected. Therefore, the spectrum of $\mathbf{A}(L(G_t))$ will be the union of the spectra of each connected component of $L(G_t)$ and Thm. S5.2 can be applied to upper-bound the smallest eigenvalue using

$$d_{\max} = \max\{d(L(g_t)) \mid g_t \in \text{components}(G_t)\}, \tag{S15}$$

where $d(L(g_t))$ is the diameter of the line graph of connected component $g_t \in G_t$ and components$(G)$ is the set of all connected component subgraphs of $G$.

## S5.1 Comparison to network corpus

Lastly, we contextualize our results above using the five networks in our corpus. For each network at each time $t$, we computed $\lambda_{\min}(\mathbf{B}_t^T \mathbf{B}_t)$ in Fig. S4 and $d_{\max}$ (Eq. (S15)) in Fig. S5. We see in Fig. S4 that the $\lambda$ gives us a good indication of the difficulty of recovering $\mathbf{E}$ from $\mathbf{N}$ and $\mathbf{B}$. For Copenhagen, Hospital, and Manufacturing Email, we observe many periods where $\lambda_{\min} = 0$, indicating that the estimation error $\left\| \mathbf{E}_t^* - \hat{\mathbf{E}}_t \right\|_2^2$ cannot be bounded. In comparison, Ant Colony and College Message should be far easier recovery problems, as $\lambda_{\min} > 0$ for most $t$. Indeed, this comports well with the overall errors we report in the main text (Fig. 5).

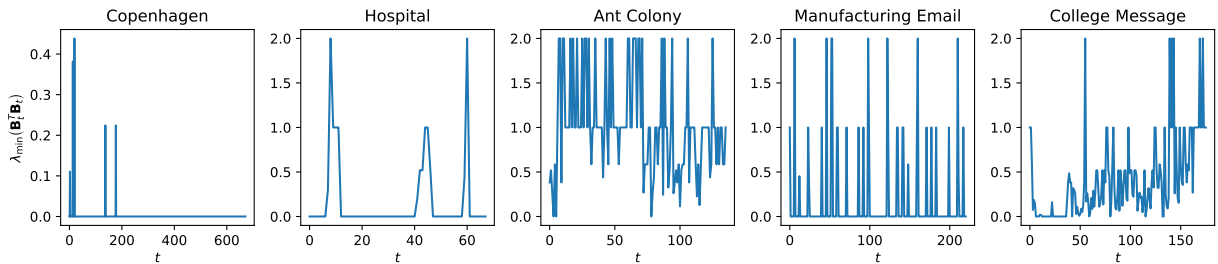

Figure S4: Minimum eigenvalue $\lambda_{\min}$ of $\mathbf{B}_t^T \mathbf{B}_t$ over time. Estimation error can only be bounded when $\lambda_{\min} > 0$. Copenhagen, Hospital, and Manufacturing Email networks are the most challenging recovery problems in our corpus, as $\lambda_{\min}$ is frequently zero. Ant Colony and College Message, in contrast, are far easier recovery problems, as both frequently have $\lambda_{\min} > 0$. In Sec. S5 we derive the graph-theoretic properties governing whether $\lambda_{\min} > 0$.

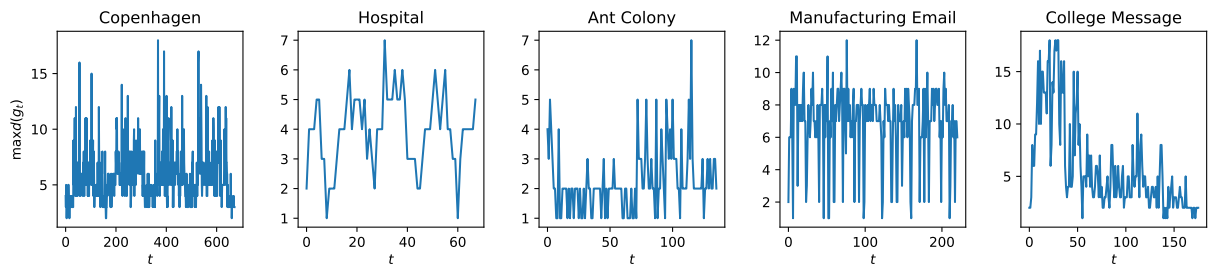

Figure S5: The maximum diameter $d$ over the connected component(s) of $G_t$ over time (see remark, Thm. S5.3).

Lastly, we support the observation in the main text that node activation fractions $n_t/n$ correlate well with estimation error, nearly as well as edge activation fractions $s_t/m$, in Table S2 (in the main text we compared activation fractions with relative estimation error). Here we again see that estimation error correlates nearly as well with $n_t/n$ as it does with $s_t/m$. While $s_t/m$ provides better correlation in all networks, $s_t$ will not be known in practice, while $n_t$ will be, and so the closeness in correlation means $n_t$ can be used as a suitable proxy for $s_t$ when trying to judge the magnitude of recovery errors.

## S5.2 Comparison to network features

Expanding up main text Sec. 4 and Fig. 5, we further report relationships between estimation error and network properties in Fig. S6. Many simple measures for an edge centrality, such as edge betweenness or various combinations of the degrees of the nodes incident to the edge, actually show weak correlation with recovery error. For the latter measures, $\min(k_i, k_j)$ tends to be most correlated with error for edge $(i, j)$—in all networks, this measure was more correlated with error than $\max(k_i, k_j)$, $(k_i + k_j)/2$, and $|k_i - k_j|$. This measure was also more correlated with error

Table S2: Correlations between $\|\mathbf{E}_t - \hat{\mathbf{E}}_t\|_2$ and the three measures. While the third measure, $n_t/n$, is the least correlated, it still gives correlation quite close in value to the other measures. This is important because any measure involving $s_t$ cannot be computed in practice. We report both Pearson correlation $r$ and Spearman correlation $\rho$.

|  | Corr. | $s_t/m$ | $\sqrt{s_t \frac{\log m}{n}}$ | $n_t/n$ |
|---|---|---|---|---|
| Copenhagen | $r$ | 0.94350 | 0.93104 | 0.84295 |
|  | $\rho$ | 0.86529 | 0.86529 | 0.79140 |
| Hospital | $r$ | 0.85576 | 0.84924 | 0.81097 |
|  | $\rho$ | 0.90346 | 0.90346 | 0.86602 |
| Ant Colony | $r$ | 0.72979 | 0.73464 | 0.70718 |
|  | $\rho$ | 0.74810 | 0.74810 | 0.73497 |
| Manufacturing Email | $r$ | 0.80367 | 0.81359 | 0.79153 |
|  | $\rho$ | 0.69007 | 0.69007 | 0.62962 |
| College Message | $r$ | 0.89974 | 0.84597 | 0.84190 |
|  | $\rho$ | 0.92486 | 0.92486 | 0.90039 |

than edge betweenness. The sparsity of the edge, the proportion of times it was active, was also correlated with error, with more active edges tending to have higher recovery errors for most networks, with College Message being a notable exception.

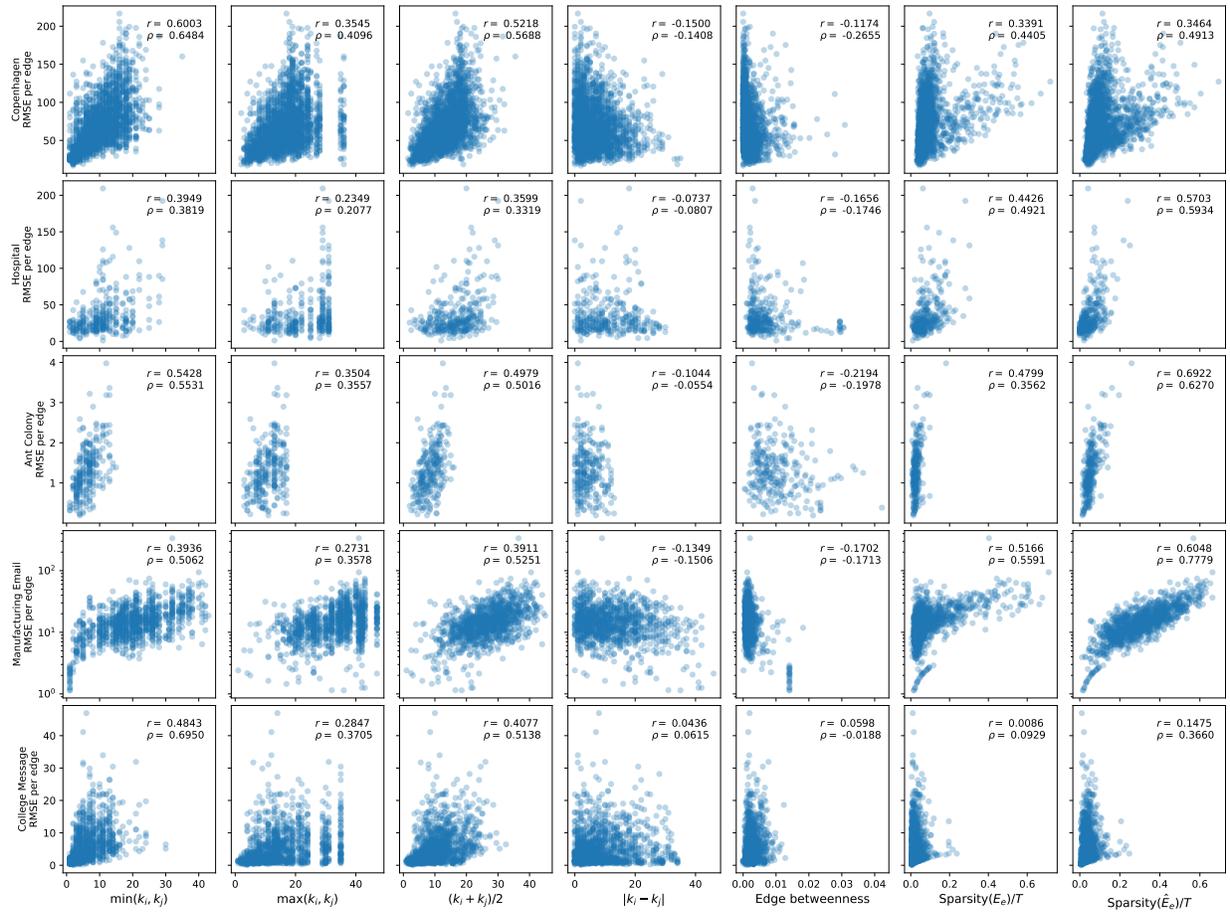

Figure S6: Empirical correlations between recovery error and edge features.


\makeatletter\@input{xx.tex}\makeatother
\end{document}